# Business Model Contributions to Bank Profit Performance: A Machine Learning Approach[*]

| Fernando Bolívar, | Miguel A. Duran, | Ana Lozano-Vivas, |
| BBVA Research (Spain) | University of Malaga (Spain) | University of Malaga (Spain) |
| fer.bolivar.garcia@gmail.com | maduran@uma.es | avivas@uma.es |

**Abstract:** This paper analyzes the relation between bank profit performance and business models. Using a machine learning–based approach, we propose a methodological strategy in which balance sheet components' contributions to profitability are the identification instruments of business models. We apply this strategy to the European Union banking system from 1997 to 2021. Our main findings indicate that the standard retail-oriented business model is the profile that performs best in terms of profitability, whereas adopting a non-specialized business profile is a strategic decision that leads to poor profitability. Additionally, our findings suggest that the effect of high capital ratios on profitability depends on the business profile. The contributions of business models to profitability decreased during the Great Recession. Although the situation showed signs of improvement afterward, the European Union banking system's ability to yield returns is still problematic in the post-crisis period, even for the best-performing group.

---

[*] The authors are very grateful to the paper's discussants and the participants at the 37th International Conference of the French Finance Association and the 6th Vietnam Symposium in Banking and Finance for helpful comments. M. A. Duran and A. Lozano-Vivas gratefully acknowledge financial support from the Spanish Ministry of Economy and Competitiveness (Project RTI2018-097620-B-I00), the European Regional Development Fund (Project FEDERJA-169), and PAIDI (Project P20_001010). F. Bolívar would also like to state that the views expressed in this paper are the authors' and do not necessarily represent those of BBVA.

# 1. Introduction

What characteristics of bank business models enhance profitability? The search for answers has been among supervisory authorities' priorities since the end of the Great Recession and will very likely become even more relevant because of the economic effects of the COVID-19 pandemic. The International Monetary Fund (2016) links poorly adapted business models to low bank profitability in advanced economies, identifying the former as an eroding agent of bank resilience over time. In a similar vein, Stephen Woulfe (2018), Head of Division at the European Central Bank, indicates that business model analysis is on authorities' agenda because it can reveal the viability of banks' strategic plans. Indeed, concern about the profitability of banks' business profiles has been especially high in the European Union (EU), as reflected by the fact that the European Central Bank launched a thematic review in 2016 to assess the relation between profitability and business models (European Central Bank 2018).

As we discuss in the literature review section, banking research has examined this relation in two broad and different ways, on the one hand, using regression analysis to explore the effect of a bank's strategic feature on profitability and, on the other hand, grouping banks into business models by means of clustering techniques.

Regression analysis of the link between business model and profitability presents some drawbacks. Regarding our research aim, the main one is that multicollinearity concerns recommend against considering all balance sheet characteristics simultaneously, thus limiting the scope of the analysis. Indeed, standard empirical practice with this type of approach explores the effect of only a reduced number of bank assets and/or liability characteristics (or just one) on profitability or focuses on a variable proxying for the type of business model (e.g., share of interest income in total income as a proxy for a nontraditional business model).

Clustering techniques remain largely immune to this drawback. However, to examine business model profitability, earlier clustering-based studies group banks according to similarities in banks' balance sheets and, then, in a second stage, compare the resulting groups' profit performance by employing descriptive statistics. Thus, instead of using profit performance as input for the clustering algorithm, these studies consider profit performance as a byproduct of the analysis.

This paper proposes a methodological strategy to explore the relation between bank profitability and business models that attempts to overcome some of the shortcomings of earlier research. In this regard, we start from the notion that a business model is a strategic choice (Hunt 1972, Caves and Porter 1977, 1978, Porter 1979). Among available options in the strategic possibility set, each bank adopts a concrete one, which is reflected in the asset and liability portfolio. Accordingly, to assess the effects of banks' actual strategic decisions, we consider the components of the complete asset and liability portfolio. Regarding the techniques structuring our methodological proposal, we resort to machine learning methods, which are not conditioned by the assumptions of standard regression analysis and could be particularly well suited to study the effect of business models' features on profitability. In particular, in line with earlier clustering-based work, we identify business models by grouping financial institutions that follow a similar strategy. However, a key contribution of our analysis is the consideration that profit performance is not just a byproduct of the grouping process, but the basis itself of the identification strategy that allows us to recognize business models. To achieve this goal, after splitting the asset and liability portfolio into its components, we compute their contributions to profitability at the observation level by means of the random forest (RF) algorithm, in combination with the tree interpreter algorithm; then, we use these contributions as the instruments of the clustering process that identifies the business profiles.

More specifically, we proceed in four stages to characterize the business models. First, using the bank portfolio's components as predictors and profitability as the response variable, we construct an ensemble of decision trees through RF (Breiman et al. 1984, Breiman 2001). In the second stage, we quantify the effects of the portfolio's components on bank profitability, that is, whether and by how much the components increase or decrease profitability. To perform this analysis, we apply a technique that considerably expands the ability to interpret RF results. In particular, this interpretation technique, the tree interpreter algorithm, allows us to compute the contributions to the response variable of the predictors chosen as splitting variables in the RF trees (Kuz'min et al. 2011, Palcewska et al. 2013, Li et al. 2019); that is, in our research, we can calculate the contributions to the profit performance of the asset–liability mix components.

The contributions provided by the tree interpreter are the instruments that we use in the third step of the empirical strategy to group banks by means of cluster analysis. Thus, the effect of the components of the business model on profit performance is the basis of our identification strategy. Once banks are clustered into groups that share similar strategic decisions in terms of contributions to profitability, we analyze their business profiles in the last stage of our proposal; in particular, we define the components of the asset and liability portfolio that characterize each business model.

The empirical exercise covers the period 1997–2021 and focuses on the banking systems of the 15 states that were members of the EU before 2004, when the EU began incorporating Eastern Europe. In line with previous studies (Ayadi et al. 2012, Martel et al. 2012), our results suggest that the standard retail-oriented business model is the best-performing profile in terms of profitability. Therefore, being traditional (in the sense of customer loans and deposits being the characteristic features of the asset and liability portfolio) seems to contribute to enhancing profitability. The worst performer in terms of returns is a non-specialized business profile, that is, a model that does not match any of the standard models identified by the banking literature. This outcome seems to indicate that adopting a strategy of specialization is better in terms of profit performance.

Additionally, we observe that the effect of a high capital ratio on profit performance depends on the business model. In particular, highly capitalized banks that adopt a non-specialized business model appear to contribute poorly to profitability, whereas banks that specialize in the standard retail-focused model and are therefore characterized by a high capital ratio have better profit performance. These findings suggest that including the bank business model as an additional variable can enrich the analysis of the potential effects of high capital ratios. In this regard, the banking literature has pointed out that high levels of capital can be an obstacle for return performance (Duran and Lozano 2015, Bank for International Settlements 2019), but our results indicate that this effect appears to characterize only non-specialized business models. Thus, when the business profile is included in the analysis, augmenting the legal capital ratio, as the Basel III Accord does, can decrease bankruptcy risk and, simultaneously, be compatible with enhancing profit performance.

Regarding the effect of the Great Recession on business models' contributions to profitability, the downturn caused a general deterioration in the profit performance of the EU banking system. After the downturn, the contributions of the two worst-performing models returned to values that, although negative, were similar to those before the crisis, but the best-performing model's total contribution deteriorated even further. In line with the EU banking authorities (Guindos 2019, Fernandez-Bollo et al. 2021), who point to low profitability as a key problem of EU banks, this outcome suggests that even the best EU performers face difficulties generating adequate levels of profits.

Our findings also indicate that business models experience some adjustment during the sample period. The least stable profile is the worst performer model, which does not match any standard strategic group. In line with previous works (Roengpitya et al. 2014, Ayadi et al. 2020), around 17% of banks, on average, migrate to alternative business models. Although the percentages of banks that migrate to worse and better business models are similar before and after the Great Recession, during the downturn the proportion of worsening banks substantially exceeded that of improving banks.

This research focuses on a key concern of supervision authorities: business model profitability. Our main contribution to the literature on this topic is twofold. First, we extend the application of the RF algorithm and tree interpreter to this area of banking analysis. The use of the latter algorithm considerably expands the set of information provided by RF's ensemble of trees; specifically, it provides information about each predictor's effect on the response variable at the data point level. This expansion of the information set is the basis of our second contribution to the literature on business model profitability. In particular, the information provided by the interpretation tool allows us to reverse the way in which previous analysis has analyzed the relation between business models and returns: instead of drawing conclusions regarding profit performance as a byproduct, the identification itself of the business models is based on the bank's asset and liability portfolio's ability to yield profits.

The remainder of the paper is structured as follows. Section 2 reviews the literature related to banks' business models and profit performance. Section 3 describes the dataset. Section 4 explains the methodological strategy. Section 5 discusses the results. Section 6 concludes the paper.

## 2. Literature review

Many studies in the banking literature have approached the relation between business profiles and profit performance by analyzing how the features of banks' strategic possibility sets affect profitability. These features can relate to internationalization strategies (Peek et al. 1999, Berger et al. 2000, DeLong 2001, Buch et al. 2004), product or funding diversification (DeYoung and Roland 2001, Stiroh 2004, Stiroh and Rumble 2006, Chiorazzo et al. 2008, Lepetit et al. 2008, Berger et al. 2010, Demirgüç-Kunt and Huizinga 2010, Brighi and Venturelli 2014, Maudos 2017), one or a few characteristics of bank balance sheets (Demirgüç-Kunt and Huizinga 1999, Goddard et al. 2004, Pasiouras and Kosmidou 2007, Athanasoglou et al. 2008, Dietrich and Wanzenried 2011,

Kanas et al. 2012), or a variable proxying for the type of business model (Altavilla et al. 2017, Detragiache et al. 2018).

Consideration of just some of the characteristics of the business profile or the reduction of the entire business model to a single variable underlies the main drawback of this approach: it does not provide an integrated view of banks' strategies regarding how they structure their balance sheets and the potential effects of these decisions on profitability.

An alternative research strategy proposes using clustering analysis to identify business profiles and then compare profit performance across models. The underpinning assumption is that banks make strategic choices that are reflected in their balance sheets and these choices permit grouping the banks into units of analysis (Hunt 1972, Caves and Porter 1977, 1978, Porter 1979). The pioneering work of Amel and Rhoades (1988) applies this type of analysis to 16 US banking markets and their balance sheet composition in 1978, 1981, and 1984. The main conclusion is that membership in the groups is relatively stable over time. Additionally, Amel and Rhoades (1998) do not find support for the widely held view that the major strategic choice for banks is between retail and wholesale banking.

Halaj and Zochowski (2009), Ayadi et al. (2011, 2012, 2019, 2020), Farnè and Vouldis (2017), Hryckiewicz and Kozlowski (2017), and Roengpitya et al. (2017) use also clustering techniques to identify the major business models in the banking sector for different sets of European countries. Overall, in contrast to the work of Amel and Rhoades (1998), these papers identify two pure business profiles, the retail- and investment-oriented models, plus two versions of these with mixed features. The empirical evidence in these studies also suggests that business profiles help explain bank profitability.

Other grouping techniques have also been used to examine the relation between banks' strategic choices and profitability. In this regard, DeSarbo and Grewal (2008) develop the concept of hybrid strategic groups, which blend the strategic recipes of more than one business profile. To identify these hybrid groups, the authors propose a multicriterion classification based on a bilinear model of overlapping classifications. Mergaerts and Vennet (2016) use factor analysis to identify business models in European countries over the period 1998–2013.

In brief, this second type of approach, an alternative to standard regression analysis, consists of two stages: the first identifies business models by cluster analysis or another grouping technique, and the second stage examines the impact of business models on profitability. A shortcoming of this two-stage approach is that profit performance is considered a byproduct of banks' strategic decisions on the balance sheet.

To overcome the drawbacks of previous studies on the relation between business models and profitability, we propose a different methodological approach. This proposal allows us, in contrast to standard regression analysis, to consider simultaneously the different components of the asset and liability portfolio. Additionally, instead of a byproduct of the analysis, profitability is the key variable for identifying bank business profiles in our proposal. Specifically, identification is based on the contributions to the profitability of banks' strategic choices.

In contrast to other areas of economic research, the use of decision tree–based techniques such as RF is relatively scarce in the banking literature.[2] Duttagupta and Cashin (2011), as well as Manasse et al. (2013), apply them to the study of banking downturns in emerging and developing countries. In a similar line of research, Ioannides et al. (2010), Alessi and Detken (2018), and Lin et al. (2022) assess the early warning signs of banking crises. Emrouznejad and Anouze (2010) combine data envelopment analysis and decision trees to carry out a bank efficiency study, whereas the focus of Bijak and Thomas (2012), Kao et al. (2012), and Pai et al. (2015) is on customer credit scoring in financial intermediation. Aparicio et al. (2018) use decision tree–based instruments to examine the relations between banks' charter value, risk taking, and supervision.

Along with extending the application of RF to a new area of banking analysis, this paper proposes using an interpretation tool, the tree interpreter algorithm, that allows us to identify the contributions of the predictor variables (in our analysis, the components of banks' asset and liability portfolio) to the response variable (bank profit performance). This tool substantially increases the amount of information that RF provides and specifically allows us to identify the types of strategic choices, using profitability as the key determinant of the grouping process.

---

[2] For instance, decision tree–based techniques have been frequently applied in the analysis of financial crises. In this sense, Ghosh and Ghosh (2002) and Frankel and Wei (2004) focus on currency crises, whereas Manasse and Roubini (2009) and Savona and Vezzoli (2015) study sovereign debt crises. Other applications analyze inactive credit card holders (Nie et al. 2011) and corporate default (Uddin et al. 2020). Jones et al. (2015, 2017) examine the performance of a set of classifiers, including RF, in predicting corporate bankruptcy and credit rating changes.

## 3. Data and components of the business model

This section describes the dataset that we use to analyze the relation between bank profit performance and business models. In addition, we define and present the summary statistics of the variables measuring the components of banks' asset–funding mix.

3.1. Data

Our sample consists of commercial banks in the 15 countries that were part of the EU until the enlargements to Eastern Europe started in 2004: specifically, Austria, Belgium, Denmark, Finland, France, Germany, Greece, Ireland, Italy, Luxembourg, the Netherlands, Portugal, Spain, Sweden, and the United Kingdom. The sample covers the period 1997–2021.

Data are obtained from the annual financial statements provided by the Fitch–IBCA Bankscope database and its successor, Moody's Analytics BankFocus database. In line with Bonin et al. (2005), we collect unconsolidated accounting data. To fully exploit the databases, we follow Mergaerts and Vennet (2016) in utilizing balance sheet data based on International Financial Reporting Standards and General Accepted Accounting Principles. We focus on commercial banks to ensure comparability across countries (Bos and Schmiedel 2007, Kontolaimou and Tsekouras, 2010, Casu et al. 2016, Badunenko et al. 2021) and to trace bank history. Variables are corrected for inflation by means of the gross domestic product deflator, with 2010 as the base year. Stock variables from the balance sheets are averaged, and flow variables from the profit and loss statements are reported year to year. Given the length of our sample period, we perform a series of cross-examinations to make the data consistent. We exclude banks missing or with clearly senseless values for basic accounting variables. Following Cetorelli and Goldberg (2012), we apply a number of filters to eliminate extreme outliers. The final sample includes 10,820 observations.

To accurately examine the relation between business models and profitability along the time interval analyzed, we split the sample into three periods: before, during, and after the Great Recession. Following Truman (2013) and Sá et al. (2016), we consider that the recession started in Europe in 2008 and extended, as the so-called sovereign debt crisis, until 2013. Accordingly, the periods that we analyze separately are 1997–2007, 2008–2013, and 2014–2021.

## 3.2. Profitability and business model framework

The response variable of our empirical exercise is bank profit performance, measured as the ratio of profit before taxes over total assets, *ROA*.

In line with previous literature, we identify true banking strategies in a portfolio context (Hryckiewicz and Kozlowski 2017); that is, we assume that banks' business strategies are reflected in their complete asset and liability structure. Indeed, most studies that analyze the relation between profitability and business models focus on the strategic choices defining the bank's asset–funding mix. Accordingly, the components of this mix feed the RF algorithm, and tree interpreter provides information about their contributions to profitability.

Our characterization of the features of banks' balance sheets somewhat follows Ayadi et al. (2011, 2012, 2019, 2020) and Roengpitya et al. (2014), but we include the complete bank asset and capital structure. These features are measured as ratios over total assets. The asset-side components are customer loans (measured as non-bank business and consumer lending), interbank lending (loans to other banks and interest-earning balances with central banks), derivative exposures (sum of positive and negative derivative transactions), and securities (securities net of derivatives). The liability components are customer deposits (non-bank business and consumer deposits), interbank borrowing (deposits from other banks), short-term funding (money market instruments and certificates of deposit), long-term funding (senior debt maturing after one year and subordinated borrowing), and equity (common equity net of intangible assets).

The literature on business models traditionally identifies, along with mixed versions, two pure models: the retail- and investment-oriented models. As Ayadi et al. (2012) note, the former concentrates on traditional activities; that is, customer loans and deposits are the most relevant balance sheet components. Banks with this model are also highly capitalized. The investment-oriented model focuses on trading activities, is predominantly funded in debt markets, and is highly leveraged. Accordingly, the main components of this business model are derivative exposures, securities, and short- and long-term borrowing, with low levels of capitalization. A general finding in this literature is that the retail-oriented model performs better than the rest of the models in terms of profitability (Ayadi et al. 2012, Martel et al. 2012).

The mixed models are the retail-diversified and wholesale models. The former is similar to the retail-oriented model, but with a more diversified funding structure. Accordingly, along with the main components of the retail-oriented, short-term funding and long-term funding represent relevant percentages of the liability side of retail-diversified banks. The main difference between the investment-oriented and wholesale models is that the latter are active in the interbank market, and therefore both interbank borrowing and lending are important for wholesale banks.

**Table 1**

Table 1 displays the descriptive statistics of bank profitability and the components of the business model.

**4. Methodological strategy**

Figure 1 synthesizes the methodological strategy that we develop to explore bank business models' profitability. After preparing the dataset according to the account in the previous section, we use RF to grow a forest of decision trees, proceed to compute business model components' contributions to profitability by means of the tree interpreter algorithm, group banks through clustering analysis, and finally analyze business models' characteristic features. The results section expounds these features, whereas the remainder of this section discusses the methodological aspects of our strategy associated with machine learning techniques.

**Figure 1**

4.1. RF

RF is an ensemble machine-learning method that builds and combines a set of individual decision trees to generate a single output, through either classification, for discrete target variables, or regression, for continuous target variables, as in our study (Breiman 2001).[3]

Decision trees are the RF model's building blocks. A single tree is formed by a set of interconnected nodes.[4] Starting from the root node, each node is linked to a couple of successor child nodes, up to the terminal ones (the leaves), which are just connected to their respective predecessors. These connections reflect successive divisions of the sample. In particular, except for the root node, which splits the entire sample, and the leaves, where the tree stops growing, each node splits the subsample that reaches it into two new portions, which are transmitted to the node's couple of child nodes. A trial-and-error process divides the subsample that a node receives. The criterion to split a node $j$ into child nodes $j_{left}$ and $j_{right}$ consists of selecting the explanatory variable (in our work, a component of the bank portfolio) and a value of this variable. This value maximizes the result from subtracting the dispersion of the response variable (profitability) at the child nodes $j_{left}$ and $j_{right}$ from the dispersion of this variable at node $j$. Thus, the set of observations at $j_{left}$ and $j_{right}$ presents maximal internal similarity and, hence, maximal comparative heterogeneity in terms of the response variable. The splitting trial-and-error process is repeated at each new node until no further meaningful segmentation is possible or a predefined stopping rule holds.

---

[3] The R package that we use to implement the RF algorithm is randomForest (https://cran.r-project.org/web/packages/randomForest/index.html).
[4] The classification and regression tree algorithm was proposed by Breiman et al. (1984). Zhang and Singer (2010) explain it in detail. Figure 3 illustrates the example of a single decision tree.

To grow the set of trees forming the forest, the RF algorithm uses two layers of randomization. The first one is obtained through bagging, which implies that individual trees are trained on subsets of the original dataset selected randomly with replacement. In particular, the algorithm draws close to 65% of the observations from the entire sample to build a tree. Feature randomness, which provides the second layer, involves the algorithm not using the entire set of predictor variables to find the best split at each node of a tree, but only a number of randomly selected ones. Both layers contribute to lower the correlation between trees and thus decrease the model's variance. Hence, in comparison to a single tree, the results become more robust to changes in the sets of predictor variables or observations, making the model more stable and allaying overfitting concerns.

If the response variable is discrete, the RF's prediction is the category voted for by the majority of trees, whereas, when the response variable is continuous, RF calculates the average of decision trees' outcomes to generate a single predicted value. Since profitability—our response variable—is continuous, the model that we develop matches the latter type. In this case, for a forest with $T$ trees generated from a training dataset $x_1, \dots, x_n$ and the corresponding values of the response variable $y_1, \dots, y_n$, if the prediction of each single tree $t$ for a new observation $x'$ is $f_t(x')$, the model's predicted value on $x'$ is $\widehat{y'} = \frac{1}{T}\sum_{t=1}^{T} f_t(x')$ (Hastie et al. 2009). In our analysis, $T$ is equal to 500; that is, we grow a 500-tree randomly generated forest.

Figure 2 focuses on whether and how the size of the trees should be controlled by a stopping rule and the number of predictors that should be evaluated at each node. In particular, this figure shows the RF model's root mean squared error (on the vertical axis) in terms of the number of predictors assessed at each node (on the horizontal axis). Each line stands for a minimum node size, that is, the minimum number of observations that any node is required to have to continue the splitting process on the corresponding branch. According to the features of the combination at which the root mean squared error is minimized in Figure 2, we choose three as the number of components used at each node to compute the best split and one observation as the minimum node size.[5] The latter feature, that is, allowing trees develop with no restraint, can make single trees overfit the data, but, at the same time, they become more dissimilar and, hence, less correlated, reducing the overall model's prediction error. Given that the bank portfolio is split into nine components, opting for three input variables being evaluated at each node is consistent with searching a third of the input variables to find the best split, which is generally accepted as a default option.

**Figure 2**

As Jones et al. (2015, 2017) point out, the RF model has considerable advantages in practical terms, making its use especially appropriate for shedding additional light on our research aim, the analysis of bank business models' profit performance. In this regard, RF is not affected by monotonic transformations of predictor variables or problems common to financial datasets, such as the presence of outliers or missing values. Additionally, it is a nonparametric method that does not require ordinary least squares assumptions or other assumptions on the data distribution. Specifically, the RF model is largely insensitive to statistical problems, such as multicollinearity. We can therefore simultaneously analyze the effect on profitability of all the components of the asset and liability portfolio. RF can also select the relevant predictor variables from a large set without being affected by the inclusion of unimportant predictors. Thus, the components of the bank portfolio used in the analysis do not have to be decided in advance. Moreover, because of the nature of the recursive splitting process characterizing RF, the selection process allows us to identify the features of the portfolio that can contribute to increase or hamper profitability, thus providing information that could help bank management enhance returns.

In relation to other machine learning techniques that share advantages with RF, the main reason to use the latter in our analysis is that, after running RF, as we explain in the following section, we can use the tree interpreter algorithm to compute the contributions of the bank portfolio's components to profitability. Such contributions indeed constitute the instrument to identify bank strategic groups and their key features. Additionally, previous comparative studies have highlighted that RF is noticeably competitive in terms of predictive accuracy, in comparison to either more conventional statistical methods or other machine learning techniques (Jones et al. 2015, 2017; Uddin et al. 2020). In line with these studies, we compare the performances of RF and alternative machine learning models, specifically, $k$-nearest neighbors, gradient boosting, and single decision trees.[6] As Table 2 suggests, among the compared methods, RF has the smallest prediction errors, as measured by the root mean squared and mean absolute errors (0.0039 and 0.0024, respectively). Given that its $R$-squared value is the highest (0.9528), RF seems to also be the model that explains the largest portion of the target variable's variation.

---

[5] Similar results are obtained if we optimize the mean absolute error or the *R*-squared value.

**Table 2**

Regarding the comparison of RF to more standard statistical techniques, such as ordinary least squares or probit/logit regressions, Jones et al. (2015, 2017) point out that there is a trade-off in the flexibility–interpretability space. More standard methods, designed to obtain a clear interpretation of the effect of a relatively reduced number of explanatory variables on the response outcome, score better in interpretability at the cost of being more rigid. By contrast, RF, along with other machine learning techniques, is more flexible: by construction, these techniques are modeled to accommodate and capture complex interactions and patterns characterizing the data. However, RF has been typically considered a black box that is difficult to interpret, since the individual analysis of forest trees can hardly be the basis of a clear, practical interpretation.

In this sense, the standard result that RF generates is the relative variable importance index, a relevance-based ranking of the predictor variables (Breiman 2001). The tree interpreter algorithm, however, is a tool that considerably enhances the interpretability of RF: it allows us to compute each predictor's contribution to the response variable at the data point level (Kuz'min et al. 2011, Palcewska et al. 2013).

---

[6] Since our aim is to explain the relation between business models and profitability during the sample period, these comparisons are made using the entire sample.

4.2. Interpretation tool

Figure 3 illustrates how the tree interpreter algorithm operates.[7] At the root node, $N0$, the response variable mean, $\bar{Y}$, is 0.3, and the predictor that generates the best split is $V_1$. Let us descend through the dashed path that starts at $N0$ and leads to the leaf that includes the data point where the response variable equals 0.6, $Y_1$. Following this path, we observe that $V_1$ contributes to modify the response variable from $N0$ to node $N1$ with a value equal to 0.2. It contributes again in the last node, $N3$, with the same value of 0.2. Predictors $V_2$ and $V_3$ produce the best splits at nodes $N1$ and $N2$, respectively, with contributions -0.4 and 0.3. If all contributions are added to $\bar{Y}$, we obtain $Y_1$. In more technical terms, $Y_1 = \bar{Y} + \sum_{i=1}^{K} cont_{V_i,Y_1}$, where $K$ is the number of predictors and $cont_{V_i,Y_1}$ is the total contribution of the $i$th predictor in the path that starts at the root node and ends in the leaf where the data point corresponding to $Y_1$ has been classified. In a forest, for this data point, a predictor's contribution is the average of its contributions across the forest trees; that is, in a forest with $T$ trees, the contribution of the $i$th predictor variable associated with the data point of $Y_1$ is $\frac{\sum_{t}^{T} cont_{V_i,Y_1,t}}{T}$.

**Figure 3**

Therefore, the tree interpreter algorithm allows us to determine the quantitative effect of the predictor variables on the response variable. In our analysis, we can determine how much each component of the asset–funding mix contributes to shift a bank's profit performance upward or downward from the average. That is, we provide information about the contribution that each predictor has on the response variable for each data point.

4.3. Identification of bank strategic groups

Cluster analysis is a technique that groups observations depending on their similarity. The level of similarity is defined in terms of a set of instrumental variables; that is, the algorithm groups observations to form clusters that are homogeneous as measured by the similarity of the instruments. Following our methodological strategy, the tree interpreter algorithm allows us to use as instruments the contributions of the bank portfolio's components to profitability. Therefore, the clusters that we obtain are based not on how similar the components of the business model are, but on how similar their effects on profitability are; that is, profitability is at the core of the strategy to identify the business models. This feature makes our approach substantially different from previous clustering-based studies of business model profitability, where profitability is not the basis of the clustering process, but a byproduct; that is, these studies use the components of the balance sheet to group banks and, then, in a second stage, compare the average profitability of the resulting groups.

---

[7] The R package used to implement this step is tree.interpreter (https://cran.r-project.org/web/packages/tree.interpreter/index.html).

To form clusters, we use the *K*-means clustering algorithm, as proposed by Hartigan and Wong (1979).[8] The purpose of the algorithm is to partition observations into *K* clusters to minimize within-cluster variance. With this aim, the *K*-means algorithm proposes a search process where the observations are iteratively reallocated to other clusters until no movement of an observation from one cluster to another will reduce the within-cluster variance. As Wu et al. (2007) point out, this clustering technique has been identified as one of the top ten algorithms in data mining and the most widely used partitional algorithm. Along with allowing for the grouping of unlabeled observations, the *K*-means method has the advantages of being efficient, relatively straightforward to implement, and easily scalable. It can also be used for clusters of different shapes and sizes and performs well with continuous data, such as those used in our research, where clustering instruments are the contributions to the profitability of bank portfolio components.

To control the sensitivity of the clustering process to the initial values of centroids, we generate 100 initial random configurations and use the best fit, that is, the set of initial centroids that minimizes the *K*-means' cost function. A drawback that *K*-means shares with other partitional clustering techniques is that the number of clusters must be previously defined and the algorithm does not provide any indication about the number that best fits our dataset. To address this question, we use the majority rule (Milligan and Cooper 1985). According to this rule, a set of indexes examines how many clusters characterize the data, with the optimal number of clusters being equal to the number recommended by the majority of the indexes.[9] Given that the aim of our clustering process is to identify bank business models, we assume that the optimal number of clusters lies between two and ten.

---

[8] We use the R package kmeans (https://www.rdocumentation.org/packages/stats/versions/3.6.2/topics/kmeans).
[9] To apply the majority rule, we implement the R package NbClust (https://cran.r-project.org /web/ packages/NbClust /index.html). For the set of indexes used to define the optimal number of clusters, see https://cran.r-project.org/web/packages/NbClust/NbClust.pdf.

## 5. Results

Our goal is to define the main strategic choices in the EU banking system. The criterion for identifying them is built around the contributions to profitability of the components of the banks' asset and liability portfolios. Thus, the relation between contributions to profitability and the strategic decisions regarding the structure of the balance sheet are at the basis of the identification process. To achieve this goal, our research follows a four-step methodological strategy.

The first step, based on the RF algorithm, is to grow a forest of decision trees, with profitability and the portfolio's components being the response and predictor variables, respectively. Table 3 shows the relative variable importance scores of the RF model's predictors. As Uddin et al. (2020) indicate, these scores capture the effect that, averaging across decision trees, predictors have on the model's performance when they are chosen as splitting variables. Thus, the scores measure the predictors' relative importance; in particular, each predictor is assigned a score equal to the percentage improvement with respect to the most important predictor, whose score is equal to 100. According to Table 3, the portfolio's component with the highest relative variable importance score is equity. Customer deposits and loans—the characteristic features of a standard retail-focused business model, along with equity—are the components with the second (60.97) and fourth (58.83) highest scores, respectively. Except for securities, which has the third highest score (60.97), the remaining characteristics of a standard wholesale business model (interbank lending and borrowing, long- and short-term funding, and derivative exposures) have the lowest scores. Therefore, the characteristic features of the retail-focused business model seem to be the predictors that, overall, contribute more to improve the accuracy of the RF model.

**Table 3**

The second step of our methodological strategy is to use the tree interpreter algorithm to compute the RF-based contributions of the portfolio's components to profitability at the bank–year level. The third step consists of feeding the $K$-means clustering algorithm with these contributions to identify the bank business profiles.

As the previous section indicates, we use the majority rule to choose the number of business models that the $K$-means algorithm identifies (Milligan and Cooper 1985).

This rule suggests that the optimal number of clusters is three; that is, in line with Ayadi et al. (2011) and Roengpitya et al. (2014), we set the number of business models equal to three. The three strategic groups of banks in our sample are named BM1, BM2, and BM3.

Before analyzing the main features of these business models, we examine the soundness of our clustering results. In particular, we perform a discriminant analysis that checks whether the contributions of the portfolio's components to profitability discriminate adequately between the business models. Panel A of Table 4 shows the Wilks lambdas and $F$-statistics of the discriminant model considering the three business models simultaneously and grouping them two by two. Additionally, Panel B reports the Wilks lambdas and $F$-statistics associated with the components' contributions in tests of the equality of group means, as well as for the three business models simultaneously and for their possible combinations in pairs.

**Table 4**

When the three business models are considered, the significance levels of the $F$-statistics suggest that the two discriminant functions estimated are significant; that is, we can reject the null hypothesis that these functions are unable to discriminate between observations. The Wilks lambda of the first discriminant function, which is the most powerful in terms of its differentiation ability,[10] is equal to 0.2574; that is, only around a quarter of the total variance in the discriminant scores is not explained by the differences among the clusters. In the cases in which we consider clusters two by two, similar results are observed, although the Wilks lambdas are slightly lower when the analysis considers only the first and second clusters (0.2073), and higher if it focuses on the first and third clusters (0.4302) or the second and fourth (0.3521).

Regarding the components' contributions to profitability, Panel B of Table 4 shows that all the $F$-statistics associated with these contributions are statistically significant, regardless of whether we consider the three business models or any two of them. Therefore, across the different definitions of the explanatory variable of our discriminant analysis, all the components' contributions seem to be relevant discriminants of the business models. According to the Wilks lambdas in Panel B, when all the business models are considered, the component whose contribution to profitability is the most relevant to discriminate between the business models is customer loans, followed by securities, interbank lending, and customer deposits. The least relevant components are (from least to most important) derivative exposures, short-term funding, equity, long-term funding, and interbank borrowing.

---

[10] In discriminant analysis, if there are enough predictor variables, the number of discriminant functions is equal to the number of categories in the target variable minus one. Accordingly, since we have three business models, the number of discriminant functions in our analysis is two. In cases such as this, the first function maximizes the between-groups differences, and the second function, which is orthogonal to the first, maximizes the same value, but controlling for the first function. Thus, the former function is the most powerful in terms of discriminating ability (Shanti 2019).

5.1. Business model features

In the last step of our methodological strategy, we identify each model's characteristic components. Table 5 reports the three business models resulting from the clustering process and their characteristic components. Panel A displays the averages of the contributions to profitability times 100 for the portfolio components, by business model. The last column of the panel shows the overall contributions of each business model. To compute these values, we sum the components' contributions at the observation level and average the results of these sums for each of the three business models.

To describe the information in Panel B of Table 5, let us focus, for the sake of simplicity, on a concrete business model, BM1, and its complementary set, S-BM1, that is, the set of observations formed by the whole sample less the observations in BM1. The row corresponding to BM1 in Panel B displays the averages in BM1 of the ratios of the portfolio's components to assets. The S-BM1 row is in light gray and shows the same averages, but in the complementary set of BM1. For instance, 0.5474 and 0.5162 are the averages of the ratios of customer loans to assets in BM1 and S-BM1, respectively.

To conclude that a component characterizes a business model, two conditions must hold: first, the value of the variable proxying for the component in the business model must be above the value in the complementary set of this model. Second, the distributions of the component in the business model and in the complementary set must be dissimilar; that is, the Wilcoxon–Mann–Whitney test must reject the null hypothesis that these distributions are drawn from the same population.

If these two requirements are satisfied, the corresponding cell in Panel B of Table 5 is in dark gray. To put it differently, if a cell in Panel B is in dark gray, the cell-associated component is characteristic of the corresponding business model. Accordingly, for the whole sample, customer loans and deposits and equity characterize BM1; interbank lending and borrowing, derivative exposures, securities, and short- and long-term funding characterize BM2; and equity characterizes BM3.

We also use the Wilcoxon–Mann–Whitney test to compare the distributions of the components in a business model to their distributions in each of the other two business models. The superscripts in Panel B of Table 5 indicate the comparisons for which this hypothesis is rejected at the 10% significance level. For instance, the fact that the superscript of derivative exposures in the business model BM2 is one (instead of 13) means that the null hypothesis is rejected when we compare the distributions of derivative exposures in BM1 and BM2, but it is not rejected when the comparison refers to BM2 and BM3. Except in the latter case (i.e., derivative exposures in BM2 vs. BM3), pairwise comparisons indicate that the distributions of any component in any pair of business models are not, in terms of the Wilcoxon–Mann–Whitney test, similar.

**Table 5**

The average contributions to profitability shown in the last column of Panel A of Table 5 are a measure that appraises the overall profit performance of business models. Hence, it allows us to sort the business models. Indeed, the business models in Table 5 are numbered accordingly, with BM1 and BM3 being the business models with the best and worst profit performance, respectively.

This profitability-based order of the business models suggests salient findings. In line with previous works (Ayadi et al. 2011, 2012, Martel et al. 2012, Roengpitya et al. 2014), the best performer, BM1, is equivalent to a standard retail-focused business model, with the main balance sheet components being customer loans on the asset side and customer deposits and equity on the liability side. The overall contribution of the balance sheet components to profit performance is positive (1.0387). Indeed, all the components contribute positively to profitability performance in BM1, regardless of whether they characterize this model. However, the characteristic components of BM1 are among those with the largest contribution.

The second-best performer, BM2, has the features of a wholesale business model. In contrast to BM1, all the balance sheet components of BM2 have negative contributions. However, four (of the six) characteristic components of this business model have the largest contributions (i.e., the lowest ones in absolute value): short-term funding (-0.0040), derivative exposures (-0.0054), interbank lending (-0.0084), and long-term funding (-0.0125). Such an outcome suggests that, although BM2 is far from having outstanding performance, most of its characteristic components have the least bad contributions.

According to Amel and Rhoades (1998), banks' strategic choices are not necessarily limited to choosing between retail and wholesale banking. In line with this finding, the worst-performing profile in our analysis, BM3, does not match any standard business model proposed by the literature. Indeed, the only characteristic component of BM3 is equity, so that it appears to be a non-specialized business model. As in BM2, the total contribution is negative (-1.2566), and all the components contribute negatively. The magnitudes of these negative contributions do not depend on the standard business model with which the components are associated. In this sense, the most negative contributions come from two components characteristic of the wholesale model (interbank lending and securities, with contributions of -0.1961 and -0.1792, respectively) and two components characteristic of the retail-focused model (customer loans and deposits, with contributions of -0.1845 and -0.1765, respectively).

Thus, for the entire sample period, EU banks that adopt specialized profiles are the best and second-best performers in terms of overall profitability, whereas the worst performer is a non-specialized business model. These results suggest that specialization seems to be a better strategic decision, as measured by profit performance, although with substantial differences between standard models: BM1, which matches the standard retail-oriented profile, is the only business model in our analysis that makes a positive overall contribution to profitability.

Equity is the only characteristic component shared by two business models, BM1 and BM3. However, equity's effect on profitability in these two models is asymmetric. Its contribution in the worst-performing model is negative (-0.0876), whereas it is positive in the best-performing one (0.2447); indeed, the contribution of equity in BM1 is, in absolute value, the largest of any component across business models. The different effect of equity on profitability in two highly capitalized business models is an outcome that can enrich the debate about the opportunity costs of increasing the regulatory capital ratio, as the Basel III Accord does. This debate shows that elevated capital ratios can diminish banks' probability of bankruptcy, but at the cost of hampering their ability to yield profits (Duran and Lozano 2015, Bank for International Settlements 2019). However, our research suggests that, if business models are included as an additional variable in this debate, enhancing profitability and reducing bankruptcy risk are not necessarily incompatible targets when the capital ratio increases. In particular, high capital ratios do not seem to be a burden for profit performance if banks specialize in a retail-oriented business model.

Our portfolio approach to the relation between profit performance and the business profile allows us to split the latter's overall contribution to the former into two: the portion linked to asset activities and that linked to financing sources. As the first two columns of Table 6 show, the contributions of the asset and liability sides are positive for BM1 (0.4147 and 0.6241, respectively), with the contribution from financing sources being greater than that generated by asset activities. Accordingly, for the best-performing model, the liability side of the portfolio seems to contribute more to enhancing profit performance. For the worst-performing model, the situation reverses: the contributions of asset and liability sides are negative (-0.6473 and -0.6093, respectively), and the effect of the former is larger. BM2 combines the features of BM1 and BM3: as in the latter case, the asset and liability sections contribute negatively to profit performance (-0.0442 and -0.1104, respectively), but the contribution of liabilities is larger, as in BM1.

**Table 6**

Overall, except for BM3, the financing structure appears to have a greater influence, either positive or negative, than asset activities on the total contribution to profitability. This finding suggests that the bank capital structure plays a determinant role not only in bank risk (Gorton and Metrick 2012), but also in bank profit performance.

## 5.2. Business models across periods

Tables 7 to 9 show the same information as in Table 5 across periods, that is, before, during, and after the Great Recession, respectively.[11] These tables indicate that the differences between the overall contributions to profitability of the best and worst performers are very similar before and after the downturn (2.1953 and 2.1764, respectively), but increase by around a 10% during the crisis (2.3967). Similar behavior is observed with respect to the second-best and worst performers (1.0034, 1.1765, and 1.0729 before, during, and after the Great Recession, respectively).

Comparison of the outcomes in the three periods also suggests that the overall contributions of BM2 and BM3 across periods follow a U-shaped path, decreasing in the crisis and increasing afterward—although maintaining negative values across periods.

---

[11] Data availability limits the possibility of using our methodological strategy to examine the relation between business models and profitability during the COVID-19 pandemic. Nevertheless, in an attempt to check whether the pandemic has modified this relation, we apply our proposal to the period 2014–2019 and compare the business models' characteristics between this period and 2014–2021. Our findings indicate there seems to be no substantial differences regarding the business models' profitability between both periods. The results for the period 2014–2019 are available upon request.

BM1 is an exception to this pattern: its overall contribution to profit performance decreases also between the Great Recession and the following period. Accordingly, although the crisis does not cause a decrease in BM1's overall contribution as sharp as those in BM2 and BM3 (-4.84%, -65.47%, and -22.49%, respectively), banks with the best-performing model are alone in suffering even further deterioration of their ability to contribute to profit performance after the Great Recession. This result is consistent with what banking authorities have identified as one of the main weaknesses of the EU banking system: the difficulties of obtaining adequate levels of return after the Great Recession (Guindos 2019, Fernandez-Bollo et al. 2021)

**Tables 7, 8, and 9**

Regarding alterations of business profiles, the three include modifications in their characteristic features. In this sense, BM1 exactly matches a standard retail-oriented model before and after the crisis, but there is a change on the asset side during the crisis: interbank lending substitutes for customer loans as a characteristic feature of the model. The second-best performer is a standard wholesale model across periods, although its investment strategy is modified after the Great Recession; in particular, securities stop being a characteristic component of BM2 in this period. The worst-performing model is the most unstable. As in the entire sample period, equity is the only characteristic component of this profile after the crisis, but interbank lending joins equity as a characteristic component before the crisis, and customer loans and long-term funding are characteristic components during the Great Recession. BM3, however, does not match a standard business model in any of the three periods.

In terms of components' contributions to profit performance, we observe that the dissimilar effect of equity in the entire sample is also present in the analysis by periods. No other component across periods or business models has a greater contribution in absolute value than equity in BM1. Such contributions range between 0.2337 and 0.2776 before and after the recession, respectively. However, the contribution of equity is negative in the other two models. The burden of this contribution is significantly heavy for the worst-performing model, especially during the recession (-0.1437). Therefore, in line with our results for the whole sample period, if high capital ratios can hamper the ability of a non-specialized business model such as BM3 to contribute to profit performance, this effect can be especially problematic during downturns.

Additionally, there are no significant differences between components' contributions in BM3 in the whole sample or by periods. All the balance sheet components have negative contributions in BM3, regardless of the period examined. The greatest negative contributions correspond to the same components as in the sample period: two components of the standard retail-oriented model (customer loans and deposits) and the other two components of the wholesale model (interbank lending and securities).

Consistent with the findings for the entire sample period, Table 6 suggests that the contributions to profitability of the asset activities and financing sources of the best (worst) performer are positive (negative) in the three periods into which we split the sample. The contributions of the asset and liability sections of the second-best–performing model are also negative in the two periods following 2008, although in the pre-crisis period the contributions of this model's assets and liabilities are positive and negative, respectively. We also observe that the overall bank liability structure has a stronger effect on profit performance than the asset side of the balance sheet in any period. The only cases in which this effect reverses are BM2 and BM3 after the recession, and in BM3 during the recession.

In addition, our analysis of how the contributions of the sides of the balance sheet evolve through periods detects asymmetric effects. The changes in the contributions of the asset and liability sections are negative between the pre-crisis and crisis periods for the three business models (except for BM2's liability section), but, in absolute terms, the effect of the asset side is larger across business models; that is, the asset side has a larger share in the deterioration that the Great Recession caused on the business models' contributions to profitability. However, from the crisis to the post-crisis period, in BM2 and BM3, liability contributions improve (52.43% and 22.17%, respectively) and asset contributions deteriorate (-28.94% and -1.73%, respectively), although the magnitude of the former substantially exceeds the latter in both business models. Therefore, capital structure seems responsible for the improvement of the overall contribution to profitability observed for BM2 and BM3 after 2013. Regarding the best-performing model, the contributions of both the asset and liability sides deteriorate from the recession to the post-crisis period, but the deterioration of the asset side is substantially greater.

5.3. Transition matrices

The analysis of how business models change across the periods surrounding the Great Recession suggests that business models undergo adjustment in time. Our last question focuses on whether banks migrate to a different business profile and whether such migration implies shifting to a model that performs better or worse in terms of its overall contribution to profitability.

To answer this question, we compute the number of banks that maintain the same business model or switch to another one between two consecutive years, $t$ and $t + 1$. Then, we calculate the percentage of banks that did not shift their business model between $t$ and $t + 1$, which is equal to the number of banks in this situation divided by the total number of sample banks present in both $t$ and $t + 1$. We also obtain the percentages of banks whose model worsens or improves between $t$ and $t + 1$, that is, those that migrated to a business model that ranks worse or better, respectively, in terms of its overall contribution to profit performance. For instance, since BM1 is the best performer, a bank worsens if it migrates from BM1 to any other business model. The numerators of these percentages of worsening and improving banks are the total numbers of banks that experience such migrations, and the denominators are the total numbers of banks present in years $t$ and $t + 1$. Once we have calculated the proportions of banks that, between any consecutive years, shift to a worse or better business model or do not migrate, we average these to obtain the corresponding percentages for the whole sample period and any of the three periods surrounding the Great Recession. Table 10 shows these mean percentages.

In line with Roengpitya et al. (2014) and Ayadi et al. (2020), the results in Table 10 indicate that the percentage of banks that migrate to a different business model is around 17%, on average, in the sample period. There is only one percentage point difference between the proportions of banks that shift to worse and better business models (9.11% and 8.12%, respectively). Accordingly, in the business model classification based on overall contribution to profit performance, the percentage of banks that drop in ranking is roughly counterbalanced by the proportion of those that rise in ranking.

**Table 10**

The comparison of the periods surrounding the Great Recession suggests that the percentages of banks that migrate to another business model are close in the three periods, ranging between 16.21% before the crisis and 18.58% during the recession. Regarding the quality of this migration, the percentage of worsening (improving) banks follows an inverted–U-shaped (U-shaped) path across periods; that is, in line with Ayadi et al. (2020), the percentage of banks migrating to a worse business model peaked during the recession (11.64% vs. 7.73% and 8.96% in the pre- and post-crisis periods, respectively), whereas the downturn presents a trough in the proportion of improving banks (6.94% vs. 8.48% and 8.47% in the pre- and post-crisis periods, respectively).

The joint analysis of business model migrations and adjustments across periods suggests, as previous works point out (Llewellyin 2013, Badunenko et al. 2021), that business profiles are not immovable but, instead, present some level of change. In our approach to the relation between business models and profitability, we find two types of variations: those that result from banks migrating to another business model and adjustments in the structure of the models.

## 6. Conclusion

The ability of business models to generate profits is on the agenda of the banking authorities, especially in the EU, and will likely become more relevant in the aftermath of the COVID-19 pandemic. We propose a new methodological strategy to identify business models. Previous works on business models have assumed the business models to be as given or have identified them using similarities in balance sheet components. In either of these cases, profit performance is a byproduct of the business model. By contrast, in our identification strategy, profitability is at the core of the definitions of banks' strategic choices. In particular, business models are the result of using the contributions of the asset and liability portfolio components to profitability as the instruments of clustering analysis.

The empirical exercise focuses on the EU banking system in 1997–2021. According to the classification of business profiles based on their overall contributions to profit performance, the best profile is a standard retail-oriented model, the second-best one is a standard wholesale model, and the worst profile, which has only equity as a characteristic component, does not match any standard model. Therefore, specialization seems to be a strategy that results in banks adopting business profiles that perform better in terms of their contributions to profitability, particularly if the banks specialize in the standard retail-oriented model.

If the contributions to profit performance of the asset and liability sides are analyzed separately, we observe a dominance effect of the capital structure in the best- and second-best–performing models; that is, financing sources seem more relevant in determining the eventual sign and volume of total contributions to profitability in BM1 and BM2. Nevertheless, the asset and liability sides' relevance regarding the model's overall contribution is reversed in the case of the worst-performing model.

Overall, we present important insights into bank authorities' concerns about the relation between profitability and business models. In this regard, the Basel III Accord has introduced tougher capital regulatory requirements to enhance banking system resilience, although, as earlier literature has pointed out, high capital ratios could weaken banks' ability to generate profits. This paper adds a new dimension to this debate. According to our results, far from being a burden for profit performance, equity is the balance sheet component with the greatest positive contribution to profitability in highly capitalized banks that adopt the standard retail-oriented model. Thus, for banks with this business profile, equity can help to enhance both the profitability and resilience of banks. However, for highly capitalized banks with a non-specialized business profile, equity contributes negatively to profit performance. Indeed, the regulatory authorities should pay attention to the outcome suggesting that, generally, non-specialized profiles seem to score poorly in profit performance.

Regarding the evolution of business models across the periods surrounding the Great Recession, the overall contributions of the three models decrease due to the downturn. After the recession, contrary to BM2 and BM3, the best-performing model cannot prevent its total contribution from deteriorating even further, underlying the difficulties of the EU banking system to generate adequate levels of returns.

Although the three models experience some adjustments in their characteristic components across periods, the least stable profile is the worst-performing model. Additionally, transition matrices suggest that the percentage of banks that shift to another business model in the sample period, approximately 17%, is almost equally split between improving and worsening banks. However, the proportion of banks that move to a worse model exceeds by about five percentage points that of banks that improve their business profile during the Great Recession.

# Figure 1. Flow chart of the modeling process

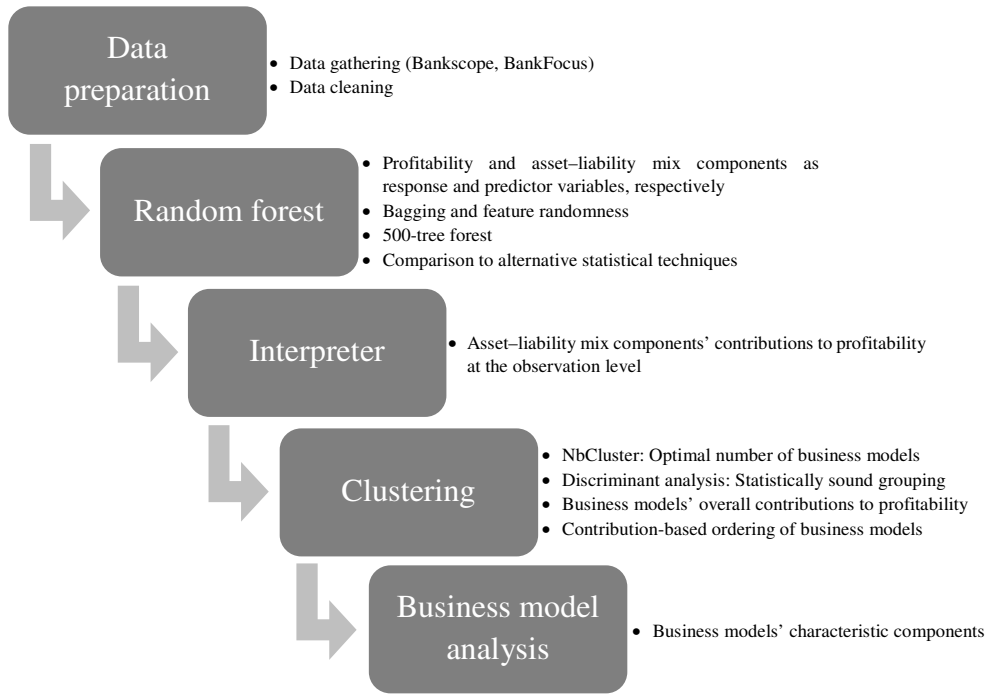

This figure synthesizes the steps of the research strategy that we follow to identify business models.

**Figure 2. The RF model: Predictor variables and node sizes**

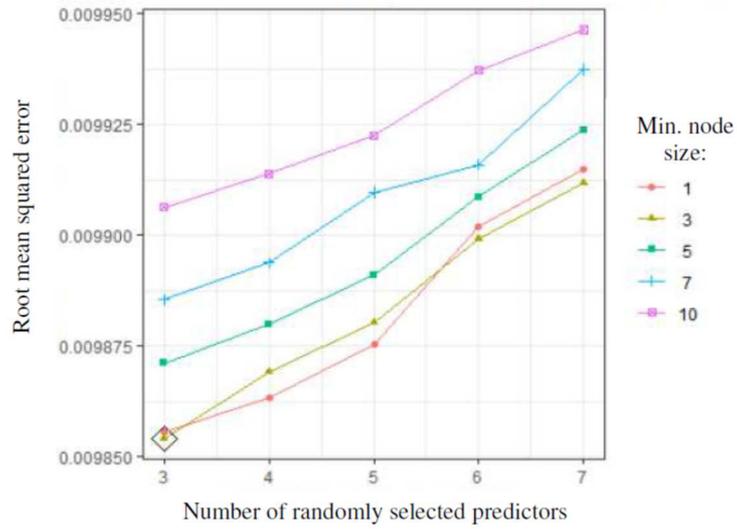

This figure shows the value of the RF model's root mean squared error (on the *y*-axis) in terms of the number of predictors that the algorithm evaluates at the tree nodes (on the *x*-axis). Each line stands for a minimum node size, from one to ten observations.

**Figure 3. Decision trees and the interpretation tool**

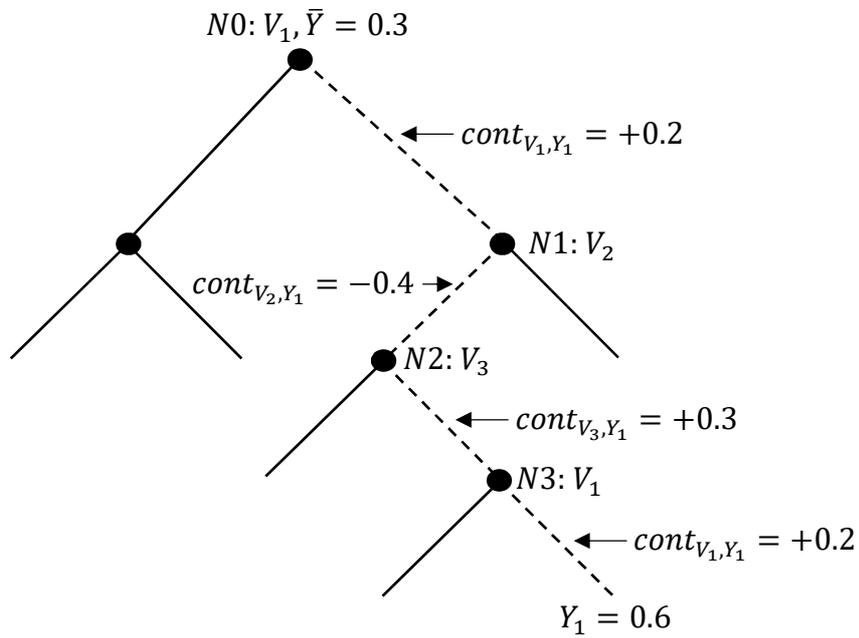

This figure illustrates the information that the tree interpreter algorithm can provide. The term $\bar{Y}$ is the sample average of the response variable, $V1$ to $V3$ are predictor variables, $Y_1$ is a concrete value of the response variable, $N0$ to $N3$ are nodes in the dashed path of the decision tree, and $cont_{V_i, Y_1}$ is the contribution to $\bar{Y}$ of the $i$th predictor at a node in the path that ends in the leaf where the data point corresponding to $Y_1$ is classified.

| Table 1. Summary statistics |||
|---|---|---|
| **Variable** | **Mean** | **St. dev.** |
| Profitability | 0.0081 | 0.0111 |
| Customer loans | 0.5227 | 0.2392 |
| Interbank lending | 0.2262 | 0.2149 |
| Derivative exposures | 0.0209 | 0.0811 |
| Securities | 0.1609 | 0.1436 |
| Customer deposits | 0.5466 | 0.2360 |
| Interbank borrowing | 0.2042 | 0.2052 |
| Short-term funding | 0.0198 | 0.0497 |
| Long-term funding | 0.0713 | 0.1016 |
| Equity | 0.0929 | 0.0638 |
| No. obs. | 10,820 ||
| This table reports descriptive statistics of the variables that measure profitability and the components of banks' asset and liability portfolio. |||

| Table 2. Machine learning models' performances ||||
| :---: | :---: | :---: | :---: |
| **Model** | **RMSE** | **MAE** | ***R*-squared** |
| RF | 0.0039 | 0.0024 | 0.9528 |
| kNN | 0.0082 | 0.0053 | 0.4963 |
| GB | 0.0101 | 0.0067 | 0.1722 |
| CART | 0.0108 | 0.0071 | 0.0569 |
| This table compares the performances of RF to three alternative machine learning models—$k$-nearest neighbors (kNN), gradient boosting (GB), and classification and regression trees (CART). The comparison is made in terms of the models' root mean squared errors (RMSE), mean absolute errors (MAE), and $R$-squared value. ||||

| Table 3. The RF model: Relative variable importance ||
|---|---|
| **Variable** | **Relative variable importance** |
| Equity | 100.00 |
| Customer deposits | 60.97 |
| Securities | 60.32 |
| Customer loans | 58.83 |
| Interbank lending | 58.07 |
| Interbank borrowing | 56.37 |
| Long-term funding | 31.92 |
| Derivative exposures | 9.81 |
| Short-term funding | 0.00 |
| No. obs. | 10,820 |
| This table reports the relative importance of the predictors in our RF empirical exercise. ||

## Table 4. Discriminant analysis

### Panel A: Discriminant functions

| All business models | | BM1–BM2 | BM1–BM3 | BM2–BM3 |
|---|---|---|---|---|
| First discriminant function | Second discriminant function | Discriminant function | Discriminant function | Discriminant function |
| 0.2574 (1166.4***) | 0.9071 (138.36***) | 0.2073 (1342.7***) | 0.4302 (1259.6***) | 0.3521 (2022.1***) |

### Panel B: Tests of the equality of group means

| Business model components | All business models | BM1–BM2 | BM1–BM3 | BM2–BM3 |
|---|---|---|---|---|
| Customer loans | 0.5150 (5092.48***) | 0.4211 (4357.40***) | 0.6623 (4368.61***) | 0.6081 (6377.66***) |
| Interbank lending | 0.5741 (4011.67***) | 0.4163 (4443.41***) | 0.6784 (4061.20***) | 0.7293 (3673.17***) |
| Derivative exposures | 0.8980 (614.22***) | 0.7468 (1074.32***) | 0.9533 (419.32***) | 0.9373 (661.59***) |
| Securities | 0.5349 (4702.59***) | 0.4293 (4213.33***) | 0.6856 (3929.44***) | 0.6366 (5650.06***) |
| Customer deposits | 0.5746 (4004.08***) | 0.4677 (3607.06***) | 0.6824 (3987.02***) | 0.6978 (4286.86***) |
| Interbank borrowing | 0.6315 (3155.81***) | 0.4502 (3869.48***) | 0.8242 (1826.85***) | 0.7146 (3952.60***) |
| Short-term funding | 0.8843 (707.90***) | 0.8474 (570.77***) | 0.9173 (772.13***) | 0.9267 (782.35***) |
| Long-term funding | 0.6558 (2838.63***) | 0.4879 (3326.52***) | 0.8357 (1684.44***) | 0.7237 (3778.11***) |
| Equity | 0.7693 (1621.94***) | 0.7382 (1123.84***) | 0.9987 (11.48***) | 0.7604 (3119.02***) |

This table reports the results from a discriminant analysis that tests whether the contributions of bank portfolio components to profitability discriminate adequately between business models. In particular, Panel A shows the Wilks lambdas and, in parentheses, the $F$-statistics of discriminant analysis, considering the three business models simultaneously and grouping them two by two. Panel B reports, for the components' contributions, the Wilks lambdas and, in parentheses, the $F$-statistics obtained in tests of the equality of group means, also for the three business models simultaneously and for their possible combinations in pairs. Statistical significance of the $F$-statistics at the 1% level is denoted by ***.

| Table 5. Business models |||||||||||
|---|---|---|---|---|---|---|---|---|---|---|
| Panel A: Mean contributions of the bank portfolio components to profitability |||||||||||
| Business model | Obs. | Customer loans | Interbank lending | Derivative exposures | Securities | Customer deposits | Interbank borrowing | Short-term funding | Long-term funding | Equity | Total contrib. |
| BM1 | 2251 | 0.1245 | 0.1087 | 0.0540 | 0.1275 | 0.1214 | 0.1261 | 0.0367 | 0.0952 | 0.2447 | 1.0387 |
| BM2 | 7649 | -0.0145 | -0.0084 | -0.0054 | -0.0160 | -0.0145 | -0.0179 | -0.0040 | -0.0125 | -0.0615 | -0.1545 |
| BM3 | 920 | -0.1845 | -0.1961 | -0.0876 | -0.1792 | -0.1765 | -0.1599 | -0.0567 | -0.1286 | -0.0876 | -1.2566 |
| Panel B: Means of the bank portfolio components |||||||||||
| Business model | Obs. | Customer loans | Interbank lending | Derivative exposures | Securities | Customer deposits | Interbank borrowing | Short-term funding | Long-term funding | Equity | |
| BM1 | 2251 | 0.5474***[23] | 0.2217**[23] | 0.0066***[23] | 0.1485***[23] | 0.5733***[23] | 0.1697***[23] | 0.0191***[23] | 0.0518***[23] | 0.1270***[23] | |
| S-BM1 | 8569 | 0.5162 | 0.2274 | 0.0247 | 0.1641 | 0.5395 | 0.2133 | 0.0200 | 0.0764 | 0.0840 | |
| BM2 | 7649 | 0.5142***[13] | 0.2304***[13] | 0.0257***[1] | 0.1659***[13] | 0.5366***[13] | 0.2162***[13] | 0.0207***[13] | 0.0779***[13] | 0.0804***[13] | |
| S-BM2 | 3171 | 0.5432 | 0.2161 | 0.0095 | 0.1487 | 0.5706 | 0.1753 | 0.0177 | 0.0554 | 0.1233 | |
| BM3 | 920 | 0.5328[12] | 0.2024***[12] | 0.0168***[1] | 0.1493**[12] | 0.5639[12] | 0.1887**[12] | 0.0142***[12] | 0.0640[12] | 0.1142***[12] | |
| S-BM3 | 9900 | 0.5218 | 0.2284 | 0.0213 | 0.1619 | 0.5449 | 0.2057 | 0.0203 | 0.0719 | 0.0910 | |

This table shows the results of applying our research proposal to the EU banking system in 1997−2021. Panel A shows the averages by business model (BM1 to BM3) of the contributions of the portfolio components to profitability. The last column of this panel refers to each business model's overall contribution to profitability. The contributions are multiplied by 100. Panel B displays the averages by business model of the ratios of the portfolio components over assets, along with the same averages for the complementary sets of business models (S-BM1 to S-BM3). To compare the distributions of each of these components in a business model and in its complementary set, we use the Wilcoxon–Mann–Whitney test. Statistical significance at the 10%, 5%, and 1% levels in this test is denoted by *, **, and ***, respectively. The same test is used to compare the distribution of each component in the business models taken two by two. The superscripts reflect the results of these comparisons. For instance, if the distribution of a component of BM1 is compared with the distribution of this component in BM2 (BM3) and the test rejects the null hypothesis at the 10% level, the value of this component in the row of BM1 has the superscript 2 (3). The cell corresponding to a given component in a business model is in dark gray if the component characterizes the business model, that is, if the average ratio of the component in the business model is above the average in the model's complementary set, and the Wilcoxon–Mann–Whitney test rejects the null hypothesis that the distributions of the component in the business model and in the complementary set are drawn from the same population at least at the 10% significance level.

| Business model | Sample period | | Before the Great Recession | | During the Great Recession | | After the Great Recession | |
|---|---|---|---|---|---|---|---|---|
| | Assets | Liabs. | Assets | Liabs. | Assets | Liabs. | Assets | Liabs. |
| **BM1** | 0.4147 | 0.6241 | 0.4343 | 0.6353 | 0.4102 | 0.6076 | 0.3389 | 0.5979 |
| **BM2** | -0.0442 | -0.1104 | 0.0127 | -0.1351 | -0.0865 | -0.1159 | -0.1115 | -0.0551 |
| **BM3** | -0.6473 | -0.6093 | -0.5157 | -0.6100 | -0.6961 | -0.6828 | -0.7082 | -0.5314 |

Table 6. Contributions of asset activities and financing sources to profit performance

This table shows the contributions to profitability of the asset and liability sides by business model for the entire sample period and the periods before, during, and after the Great Recession. The contributions are multiplied by 100.

| Table 7. Business models before the Great Recession ||||||||||||
|---|---|---|---|---|---|---|---|---|---|---|---|
| Panel A: Mean contributions of the bank portfolio components to profitability ||||||||||||
| Business model | Obs. | Customer loans | Interbank lending | Derivative exposures | Securities | Customer deposits | Interbank borrowing | Short-term funding | Long-term funding | Equity | Total contrib. |
| BM1 | 1461 | 0.1315 | 0.1187 | 0.0590 | 0.1252 | 0.1273 | 0.1268 | 0.0529 | 0.0946 | 0.2337 | 1.0696 |
| BM2 | 3735 | -0.0127 | 0.0031 | 0.0357 | -0.0134 | -0.0068 | -0.0198 | 0.0035 | -0.0143 | -0.0976 | -0.1223 |
| BM3 | 270 | -0.1998 | -0.1671 | 0.0263 | -0.1751 | -0.1776 | -0.1567 | -0.0733 | -0.1324 | -0.0700 | -1.1257 |
| Panel B: Means of the bank portfolio components ||||||||||||
| Business model | Obs. | Customer loans | Interbank lending | Derivative exposures | Securities | Customer deposits | Interbank borrowing | Short-term funding | Long-term funding | Equity | |
| BM1 | 1461 | 0.5518***23 | 0.2335***23 | 0.0021***23 | 0.1433***2 | 0.5581***23 | 0.1796***23 | 0.0247**2 | 0.0561***23 | 0.1223***2 | |
| S-BM1 | 4005 | 0.4878 | 0.2720 | 0.0084 | 0.1714 | 0.5041 | 0.2423 | 0.0286 | 0.0850 | 0.0773 | |
| BM2 | 3735 | 0.4884***1 | 0.2699***13 | 0.0090***13 | 0.1732***13 | 0.5035***1 | 0.2442***13 | 0.0289**1 | 0.0862***13 | 0.0738***13 | |
| S-BM2 | 1731 | 0.5405 | 0.2441 | 0.0019 | 0.1438 | 0.5510 | 0.1854 | 0.0246 | 0.0580 | 0.1229 | |
| BM3 | 270 | 0.47931 | 0.3016***12 | 0.0008***12 | 0.1464**2 | 0.51291 | 0.2169**12 | 0.0237 | 0.068412 | 0.1261***2 | |
| S-BM3 | 5196 | 0.5062 | 0.2596 | 0.0070 | 0.1648 | 0.5188 | 0.2260 | 0.0277 | 0.0778 | 0.0875 | |
| For the period before the Great Recession, this table shows results similar to those in Table 5. ||||||||||||

| Table 8. Business models during the Great Recession |||||||||||
|---|---|---|---|---|---|---|---|---|---|---|
| Panel A: Mean contributions of the bank portfolio components to profitability |||||||||||
| Business model | Obs. | Customer loans | Interbank lending | Derivative exposures | Securities | Customer deposits | Interbank borrowing | Short-term funding | Long-term funding | Equity | Total contrib. |
| BM1 | 438 | 0.1194 | 0.1023 | 0.0475 | 0.1410 | 0.1237 | 0.1215 | 0.0107 | 0.0960 | 0.2557 | 1.0178 |
| BM2 | 2040 | -0.0147 | -0.0146 | -0.0384 | -0.0188 | -0.0204 | -0.0190 | -0.0113 | -0.0160 | -0.0492 | -0.2024 |
| BM3 | 333 | -0.1873 | -0.2109 | -0.1209 | -0.1771 | -0.1746 | -0.1695 | -0.0600 | -0.1348 | -0.1437 | -1.3789 |
| Panel B: Means of the bank portfolio components |||||||||||
| Business model | Obs. | Customer loans | Interbank lending | Derivative exposures | Securities | Customer deposits | Interbank borrowing | Short-term funding | Long-term funding | Equity | |
| BM1 | 438 | 0.5203³ | 0.2410***²³ | 0.0186***²³ | 0.1482***²³ | 0.5722***²³ | 0.1714***²³ | 0.0119***²³ | 0.0510***²³ | 0.1343***²³ | |
| S-BM1 | 2373 | 0.5416 | 0.2117 | 0.0395 | 0.1567 | 0.5462 | 0.1990 | 0.0159 | 0.0832 | 0.0853 | |
| BM2 | 2040 | 0.5351³ | 0.2188**¹³ | 0.0421***¹ | 0.1575***¹ | 0.5446***¹ | 0.2008***¹ | 0.0161***¹ | 0.0834***¹ | 0.0826***¹³ | |
| S-BM2 | 771 | 0.5466 | 0.2094 | 0.0209 | 0.1498 | 0.5652 | 0.1786 | 0.0131 | 0.0645 | 0.1202 | |
| BM3 | 333 | 0.5812***¹² | 0.1678***¹² | 0.0238¹ | 0.1518¹ | 0.5561¹ | 0.1880¹ | 0.0146¹ | 0.0824**¹ | 0.1016*¹² | |
| S-BM3 | 2478 | 0.5325 | 0.2227 | 0.0380 | 0.1558 | 0.5495 | 0.1956 | 0.0154 | 0.0777 | 0.0918 | |

This table shows results similar to those in Table 5 for the Great Recession.

| Table 9. Business models after the Great Recession ||||||||||||
|---|---|---|---|---|---|---|---|---|---|---|---|
| Panel A: Mean contributions of the bank portfolio components to profitability ||||||||||||
| Business model | Obs. | Customer loans | Interbank lending | Derivative exposures | Securities | Customer deposits | Interbank borrowing | Short-term funding | Long-term funding | Equity | Total contrib. ||
| BM1 | 352 | 0.1019 | 0.0752 | 0.0414 | 0.1204 | 0.0940 | 0.1291 | 0.0018 | 0.0965 | 0.2766 | 0.9368 ||
| BM2 | 1874 | -0.0177 | -0.0246 | -0.0512 | -0.0180 | -0.0234 | -0.0128 | -0.0109 | -0.0052 | -0.0028 | -0.1667 ||
| BM3 | 317 | -0.1685 | -0.2054 | -0.1495 | -0.1848 | -0.1774 | -0.1525 | -0.0392 | -0.1187 | -0.0436 | -1.2396 ||
| Panel B: Means of the bank portfolio components ||||||||||||
| Business model | Obs. | Customer loans | Interbank lending | Derivative exposures | Securities | Customer deposits | Interbank borrowing | Short-term funding | Long-term funding | Equity |||
| BM1 | 352 | 0.5628*[23] | 0.1486***[2] | 0.0099***[23] | 0.1702 | 0.6382***[23] | 0.1267***[23] | 0.0047***[2] | 0.0350***[23] | 0.1376***[23] |||
| S-BM1 | 2191 | 0.5407 | 0.1629 | 0.0385 | 0.1588 | 0.5970 | 0.1756 | 0.0086 | 0.0531 | 0.0947 |||
| BM2 | 1874 | 0.5430[1] | 0.1644***[13] | 0.0411***[1] | 0.1604 | 0.5939***[1] | 0.1773***[1] | 0.0092***[13] | 0.0552***[1] | 0.0909***[13] |||
| S-BM2 | 669 | 0.5461 | 0.1513 | 0.0161 | 0.1602 | 0.6274 | 0.1451 | 0.0051 | 0.0379 | 0.1279 |||
| BM3 | 317 | 0.5275[1] | 0.1542[2] | 0.0231[1] | 0.1491 | 0.6155[1] | 0.1655[1] | 0.0056***[2] | 0.0411[1] | 0.1172***[12] |||
| S-BM3 | 2226 | 0.5461 | 0.1619 | 0.0362 | 0.1620 | 0.6009 | 0.1693 | 0.0085 | 0.0520 | 0.0983 |||

This table shows results similar to those in Table 5 for the period after the Great Recession.

| Table 10. Business model migration (in percentages) | | | | |
|---|---|---|---|---|
|  | **Sample period** | **Before the Great Recession** | **During the Great Recession** | **After the Great Recession** |
| **Worse** | 9.11 | 7.73 | 11.64 | 8.96 |
| **Equal** | 82.77 | 83.79 | 81.41 | 82.57 |
| **Better** | 8.12 | 8.48 | 6.94 | 8.47 |
| For the entire sample and the periods before, during, and after the Great Recession, this table shows the average percentages of banks that maintained the same business model (the row marked *equal*) or moved to another business model that was worse or better in terms of its overall contribution to profitability (the rows labeled *worse* and *better*, respectively). | | | | |